\RequirePackage{fix-cm}
\documentclass[smallextended]{svjour3}       % onecolumn (second format)
\smartqed  % flush right qed marks, e.g. at end of proof
\usepackage{graphicx}
\begin{document}

\title{A measurement of two-photon exchange in Super-Rosenbluth separations with positron beams}

\titlerunning{TPE exchange in e$^+$-p Super-Rosenbluth separations}    % if too long for running head

\author{John R Arrington         \and
        Mikhail Yurov
}

\institute{J. Arrington \at
              Lawrence Berkeley National Laboratory, Berkeley, CA 94720, USA \email{JArrington@lbl.gov}
           \and
           M. Yurov \at
              Los Alamos National Laboratory, Los Alamos, NM, 87545, USA \\
}

\date{Received: date / Accepted: date}
% The correct dates will be entered by the editor

\maketitle

\begin{abstract}

The proton electric and magnetic form factors, $G_E$ and $G_M$, are intrinsically connected to the spatial distribution of charge and magnetization in the proton. For decades, Rosenbluth separation measurements of the angular dependence of elastic e$^-$-p scattering were used to extract $G_E$ and $G_M$. More recently, polarized electron scattering measurements, aiming to improve the precision of $G_E$ extractions, showed significant disagreement with Rosenbluth measurements at large momentum transfers ($Q^2$). This discrepancy is generally attributed to neglected two-photon exchange (TPE) corrections.

At larger $Q^2$ values, a new `Super-Rosenbluth' technique was used to improve the precision of the Rosenbluth extraction, allowing for a better quantification of the discrepancy, while comparisons of e$^+$-p and e$^-$-p scattering indicated the presence of TPE corrections, but at $Q^2$ values below where a clear discrepancy is observed.
In this work, we demonstrate the significant benefits to combining the Super-Rosenbluth technique with positron beam measurements. This approach provides a greater kinematic reach and is insensitive to some of the key systematic uncertainties in previous positron measurements.

\end{abstract}

\section{Nucleon form factors measurements} \label{sec:intro}

The proton electromagnetic form factors provide unique insight into the proton's spatial structure, encoding the radial distribution of its charge and magnetization~\cite{Kelly:2002if,Miller:2018ybm}. In the one-photon exchange approximation, the proton form factors can be related to the reduced cross section for
electron-proton (e$^-$-p) or positron-proton (e$^+$-p) elastic scattering~\cite{Perdrisat:2006hj,Arrington:2006zm},
\begin{equation}
\sigma_R(Q^2,\varepsilon) = \tau G_M^2(Q^2) + \varepsilon G_E^2(Q^2) ,
\end{equation}
where $G_E(Q^2)$ and $G_M(Q^2)$ are the charge and magnetic form factors of the proton, $Q^2$ is the four-momentum transfer squared, $\tau=Q^2/4M_p^2$, $M_p$ is the mass of the proton, $\varepsilon = 1/\left[1 \,+\, 2(1+\tau) \tan^2(\theta_e/2)\right]$ is the virtual photon polarization parameter, and $\theta_e$ is the electron scattering angle. The form factors $G_E$ and $G_M$ are normalized at $Q^2=0$ to the proton's charge (1) and magnetic moment ($\mu_p)$ in units of elementary charge and nuclear magneton, respectively.

At fixed $Q^2$, $\sigma_R$ depends linearly on $\varepsilon$. By keeping $Q^2$ fixed but making measurements at multiple $\varepsilon$ values (by varying the beam energy and electron scattering angle) one can map out the $\varepsilon$ dependence of the reduced cross section. A linear fit in $\varepsilon$ allows for the extraction of $G_M(Q^2)$ from the value of the fit at $\varepsilon=0$ and $G_E(Q^2)$ from the slope. This method of extracting the form factors is known as a Rosenbluth separation~\cite{Rosenbluth:1950yq} or Longitudinal-Transverse (LT) separation.

Measurements of the elastic e$^-$-p cross section from the 1960s through the 1990s showed that the form factors approximately followed the dipole form,
\begin{equation}
    G_E(Q^2) \approx G_M(Q^2)/\mu_p^2 \approx G_D(Q^2) = 1/(1+Q^2/0.71)^2 ,
\end{equation}
with $Q^2$ in units of GeV$^2$. Therefore, even for $\varepsilon=1$, the contribution of the charge form factor is suppressed relative to the magnetic form factor by a factor of $1/(\tau \mu_p^2) \approx 0.5/Q^2$. At $Q^2 \approx 1$~GeV$^2$, $G_E$ contributes at most 30\% to the cross section, and its contribution decreases as $1/Q^2$ at larger $Q^2$ values. Because of this, direct Rosenbluth separations are severely limited in precision above $Q^2=3$-4~GeV$^2$~\cite{Arrington:2003df}. In addition, the extraction of $G_E$ is very sensitive to any experimental corrections that vary with $\varepsilon$, which was a particular problem for analyses that combined different experiments to obtain data covering a range of $\varepsilon$.

Polarization measurements provided a way to overcome the limitations of the Rosenbluth technique at larger $Q^2$ values~\cite{Akhiezer:1957aa,Arnold:1980zj}. Measurements utilizing polarized electron beams in combination with polarized targets or recoil polarization measurements, referred to collectively as PT measurements, are sensitive to the combination $R_p=\mu_p G_E/G_M$, rather than the squares of the individual form factors. As such, polarization measurements of $R_p$ combined with Rosenbluth separations, which yield precise values for $G_M$ at large $Q^2$, were expected to allow for reliable measurements of both form factors over a wide range of $Q^2$.

\begin{figure}
    \centering
    \includegraphics[width=0.6\textwidth,height=0.4\textwidth,trim={40mm 140mm 15mm 40mm}, clip]{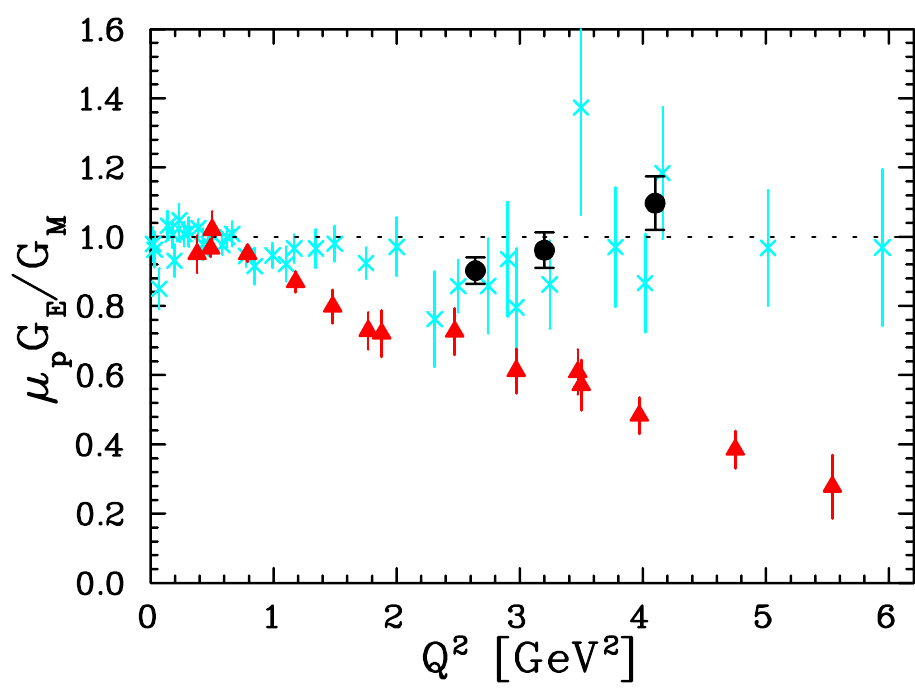}
    \caption{Measurements of $R_p=\mu_p G_E/G_M$ for the proton. Red triangles are polarization measurements~\cite{Jones:1999rz,Gayou:2001qd}, Cyan crosses are Rosenbluth extractions~\cite{Arrington:2003qk}, and black circles are the Super-Rosenbluth measurements~\cite{Qattan:2004ht}. Figure adapted from~\cite{Qattan:2004ht}. }
    \label{fig:LTPT}
\end{figure}

The first precision measurements of the form factors utilizing polarization observables~\cite{Jones:1999rz} showed a significant deviation from the Rosenbluth observation of $R_p \approx 1$, with $R_p$ decreasing approximately linearly with $Q^2$. This dramatic discrepancy led to reexaminations of previous Rosenbluth extractions~\cite{Arrington:2003df,Arrington:2003qk} as well as new high-$Q^2$ Rosenbluth extractions~\cite{Qattan:2004ht,Christy:2004rc}, and new polarization measurements~\cite{Punjabi:2005wq,Puckett:2010ac,Puckett:2011xg,Puckett:2017flj}. These efforts demonstrated that there was a clear inconsistency, illustrated in Figure~\ref{fig:LTPT}, between the form factors extracted using these two techniques up to $Q^2 = 6$~GeV$^2$, and that the discrepancy was above what could be explained by the experimental and theoretical uncertainties quoted by the measurements~\cite{Arrington:2003df,Arrington:2003qk,Qattan:2004ht}.

\begin{figure}
    \centering
    \includegraphics[width=0.96\textwidth]{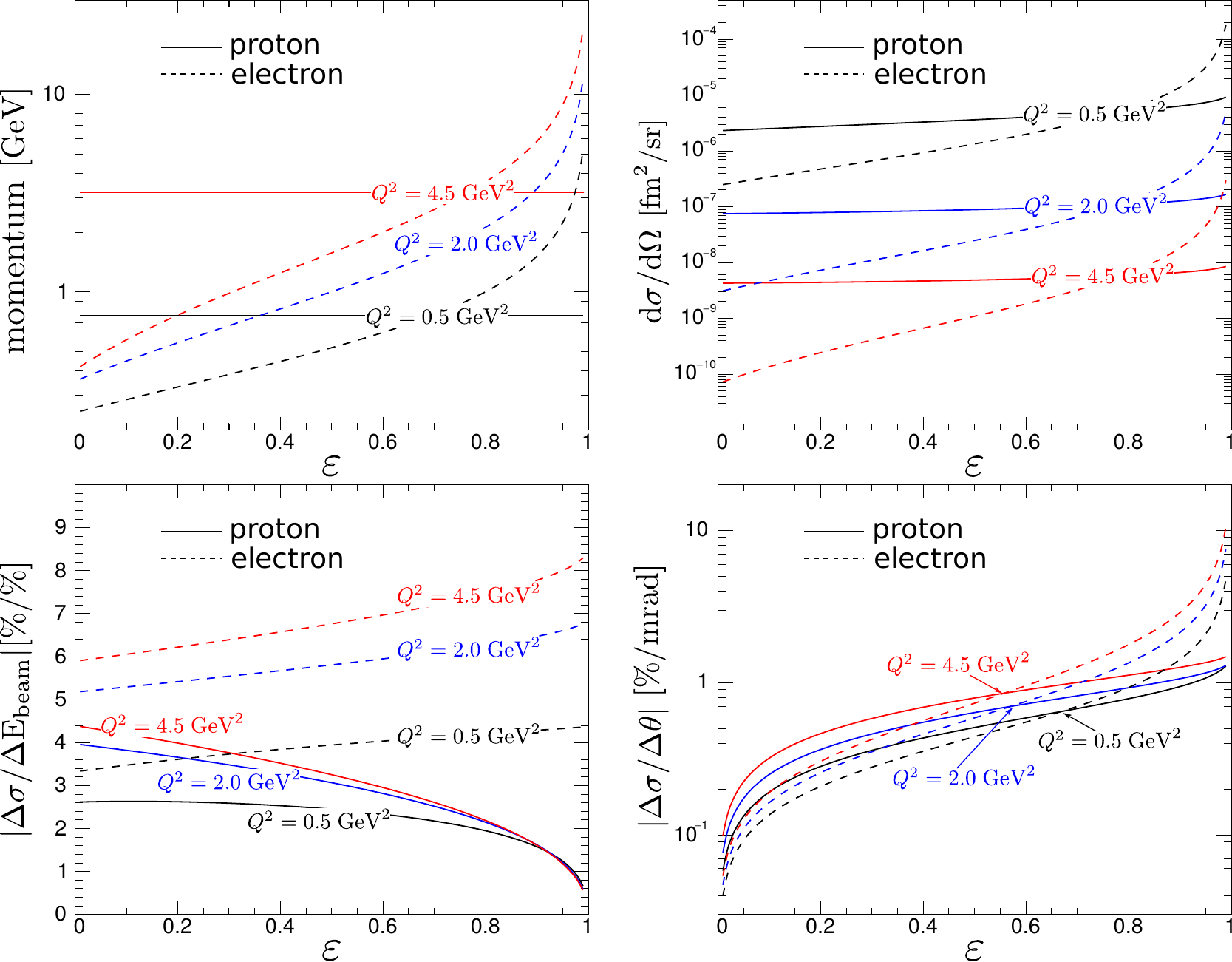}
%    \includegraphics[width=0.46\textwidth,height=0.33\textwidth,trim={3mm 2mm 5mm 3mm},clip]{f_eSR_pSR_compare_mom.pdf} \hspace{0.1cm}
%    \includegraphics[width=0.46\textwidth,height=0.33\textwidth,trim={3mm 2mm 5mm 3mm},clip]{f_eSR_pSR_compare_xsec.pdf}
%    \vspace{0.1cm}
%    \includegraphics[width=0.46\textwidth,height=0.33\textwidth,trim={3mm 2mm 5mm 3mm},clip]{f_eSR_pSR_compare_ebeam.pdf} \hspace{0.1cm}
%    \includegraphics[width=0.46\textwidth,height=0.33\textwidth,trim={3mm 2mm 5mm 3mm},clip]{f_eSR_pSR_compare_ang.pdf} \\
    \caption{Comparison of electron and proton detection in elastic e$^-$-p scattering for a range of $Q^2$ values. The top left panel shows the detected particle momentum vs $\varepsilon$, showing that proton detection at fixed $Q^2$ is insensitive to momentum-dependent detector response. The top right panel shows the cross section, which varies very little with $\varepsilon$ and is significantly higher for proton detection in the low-$\varepsilon$ region. The bottom left (right) panel show the cross section sensitivity to uncertainty in beam energy (particle angle).  The sensitivity to beam energy is lower for proton detection in all kinematics, and the sensitivity to angle dramatically reduced for large $\varepsilon$ measurements.}
    \label{fig:SRplots}
\end{figure}

An important work in confirming and quantifying this discrepancy was the so-called ``Super-Rosenbluth'' experiment~\cite{Qattan:2004ht}. The Super-Rosenbluth (SR) separation is identical to conventional Rosenbluth measurements except for the fact that the struck proton, rather than the scattered electron, is detected. Proton detection provides a large number of benefits when interested in the ratio $R_p$~\cite{Qattan:2004ht,Qattan:2005zd}, which depends only on the relative $\varepsilon$ dependence of the reduced cross section.  Figure~\ref{fig:SRplots} illustrates some of the advantages of the SR technique. At fixed $Q^2$, the proton momentum is constant, so all momentum-dependent corrections cancel out when examining the $\varepsilon$ dependence. The cross section for proton detection has a dramatically smaller $\varepsilon$ dependence compared to electron detection, making the measurements less sensitive to rate-dependent effects and removing the need to increase beam current (and thus target heating corrections) as the cross section decreases. The cross section is generally much less sensitive to the knowledge of the beam energy and angle of the detected particle. Finally, the cross section is insensitive to radiative corrections where the scattered (undetected) electron radiates a photon, reducing both the size and $\varepsilon$-dependence of the radiative corrections~\cite{Qattan:2005zd}. Because of these advantages, the SR measurements allowed for an extraction of $R_p$ with precision comparable to the polarization measurements. These new, precise results were consistent with conventional Rosenbluth extractions (Fig.~\ref{fig:LTPT}) and were significantly less sensitive to experimental systematics. This helped to rule out several potential experimental corrections that may have caused the discrepancy, provided a test of the standard radiative correction procedures, and gave a better quantitative measurement of the discrepancy.

\section{Two-photon exchange corrections}\label{TPE}

While experimental efforts were focusing on confirming the discrepancy, the contribution of TPE diagrams were being examined as a potential explanation for the discrepancy~\cite{Guichon:2003qm,Blunden:2003sp}. TPE corrections are expected to be small for both cross section and polarization measurements, but it was demonstrated that a relatively small, $\varepsilon$-dependent TPE correction could significantly impact the extraction of $G_E$ from high-$Q^2$ Rosenbluth measurements. At large $Q^2$, the $\varepsilon$ dependence arising from $G_E$ is only 5-10\% of the reduced cross section, making even a percent-level correction significant if it modifies the $\varepsilon$ dependence. In addition to changing the Rosenbluth slope, TPE contributions can also cause the reduced cross section to deviate from the linear behavior expected in the Born approximation~\cite{Blunden:2003sp,Chen:2004tw,Arrington:2011dn}. 

As noted above, the Super-Rosenbluth experiment~\cite{Qattan:2004ht} was instrumental in confirming that the discrepancy was significant and could not be easily explained by experimental uncertainties. The experiment gained additional importance in the context of TPE as a likely source of the discrepancy. The high-precision extraction of $R_p$ allowed for the best quantification of the LT-PT discrepancy and thus the size of the linear TPE contributions. In addition, it provided significantly improved tests for non-linear contributions from TPE exchange~\cite{Tvaskis:2005ex}. The quantification of the LT-PT discrepancy, combined with the limits on non-linear contributions, allows for an extraction of the form factors~\cite{Arrington:2007ux,Bernauer:2013tpr,Ye:2017gyb} from combined analysis of polarization data and cross section (with calculated and/or phenomenological TPE corrections). Under the assumptions of these combined fits, the TPE corrections do not dominate the uncertainties in the extracted form factors. However, this relies on the assumption that TPE contributions fully explain the observed discrepancy, and so far there is no direct observation of TPE for $Q^2 \geq 2$~GeV$^2$. Without knowing that TPE fully resolve the discrepancy, we cannot be certain that these extractions are reliable.

More direct tests were made by comparing positron-proton and electron-proton scattering, where the TPE contributions change sign. A global analysis of earlier measurements~\cite{Arrington:2003ck} showed evidence for TPE contributions with an $\varepsilon$ dependence consistent with the observed discrepancy, but the data showing non-zero TPE contributions were limited to $Q^2$ values below 1~GeV$^2$. In addition, new experiments were proposed and carried out to study TPE in e$^-$-p and e$^+$-p scattering~\cite{Adikaram:2014ykv,Rimal:2016toz,Rachek:2014fam,Henderson:2016dea}. These experiments confirmed the presence of TPE contributions up to $Q^2 \approx 1.5$~GeV$^2$ and were in qualitative agreement with TPE calculations~\cite{Blunden:2005ew,Zhou:2014xka}, but did not extend to the $Q^2$ region where a clear discrepancy was observed and lacked sufficient precision to look for non-linear contributions in $\varepsilon$.

At the present time, while there are significant indications that TPE corrections are responsible for the form factor discrepancy, there is no direct confirmation. Direct extractions of TPE from e$^+$-p/e$^-$-p cross section ratios indicate the presence of TPE corrections, but do not extend to $Q^2$ values where a large discrepancy is observed. Comparisons of LT and PT measurements~\cite{Arrington:2003df,Qattan:2005zd}, including a recent result with improved radiative correction procedures~\cite{Gramolin:2016hjt}, show indications of a discrepancy at the 2$\sigma$ level up to $Q^2 \approx 6$~GeV$^2$ (and only one sigma at 8~GeV$^2$)~\cite{GMP12}, but cannot identify the source of the discrepancy. Additional understanding is required to have reliable extractions of the form factors and to validate calculations of these corrections for other electron scattering observables~\cite{Arrington:2003qk,Arrington:2006hm,Blunden:2009dm,Arrington:2011dn,Arrington:2011kb,Blunden:2012ty,Hall:2013loa,Hall:2015loa,Afanasev:2017gsk}:
\begin{itemize}
    \item Confirmation of TPE as the source of the discrepancy above 1.5~GeV$^2$;
    \item Better constraints on the size of TPE from improved Rosenbluth separations for $Q^2>2$-3~GeV$^2$;
    \item Improved constraints on non-linear contributions for all $Q^2$ values.
\end{itemize}
Additional data exist that can help address the second of these questions. Data from high-$Q^2$ form factor measurements~\cite{GMP12} can extend the $Q^2$ range of the LT-PT comparisons above 6~GeV$^2$, while additional Super-Rosenbluth measurements up to 6~GeV$^2$~\cite{SR2} will improve the constraints on TPE in this region as well as improve measurements of (or constraints on) non-linear contributions. However, we still have no direct evidence that the source of the discrepancy in the region of significant LT-PT discrepancy is entirely due to TPE. Without a direct demonstration that TPE fully explains the discrepancy, the assumptions currently being made to extract the form factors could yield incorrect results. In addition, testing TPE calculations in elastic e$^-$-p and e$^+$-p scattering at modest-to-high $Q^2$ values will give us confidence that such calculations can be applied to other observables. 

In the remainder of this paper, we lay out a proposal to combine the benefits of the Super-Rosenbluth technique with the direct sensitivity of e$^+$-p/e$^-$-p cross section comparisons. Several of the benefits provided by the SR measurement address challenges in the direct e$^+$-p/e$^-$-p cross section comparisons, allowing for a direct confirmation of the TPE hypothesis, as well as providing significantly improved constraints on both the size and non-linear contributions of TPE compared to electron Super-Rosenbluth measurements or conventional direct comparisons of e$^+$-p and e$^-$-p scattering. The discussion here expands on the ideas presented in Refs.~\cite{yurov2017,Accardi:2020swt}.

\section{Super-Rosenbluth measurements with positron beams}\label{sec:SR-positron}

While e$^+$-p/e$^-$-p cross section comparisons provide the most direct measurement of TPE, they are extremely challenging in practice. Both collider measurements and fixed target experiments utilizing secondary positron beams provide modest luminosity, making it challenging to reach larger $Q^2$ and low $\varepsilon$ values where the cross section is low, but where TPE contributions are expected to be largest. In some cases, there are different corrections and systematic uncertainties associated with e$^+$ and e$^-$ beams, limiting the sensitivity even where sufficient statistics can be collected. Differences between positron and electron running conditions can also limit the precision of the measurements in cases where it is not possible to change between e$^-$ and e$^+$ beams quickly, which can lead to different run conditions for the positron and electron data. Finally, measurements utilizing a fixed beam energy do not allow for a direct extraction at fixed $Q^2$ at different angles, and thus cannot directly measure the $\varepsilon$ dependence at fixed $Q^2$.

Several of these limitations are reduced or eliminated when using the Super-Rosenbluth technique. As shown in Fig.~\ref{fig:SRplots}, the cross section for proton detection is significantly higher than for electron detection at low $\varepsilon$, where TPE are largest, offsetting the low luminosity and extending the kinematic reach of the measurements. This also allows for measurements to be taken at a fixed beam current, avoiding significant changes in rate-dependent corrections or target heating effects that can be significant when using high beam currents for measurements at low-cross section kinematics. By focusing on the $\varepsilon$ dependence, the extraction of $R_p = \mu_p G_E/G_M$ comes from comparison of positron cross sections measured at different kinematics. Only after the extraction of $R_p$ from the positron measurements do we compare it to electron results, making the result significantly less sensitive to differences between electron and positron beam characteristics, and eliminating the need for careful comparisons of beam quality or frequent changes between electron and positron beams to account for potential long-term drifts in the detector response. 

\begin{figure}
    \centering
    \includegraphics[width=0.96\textwidth]{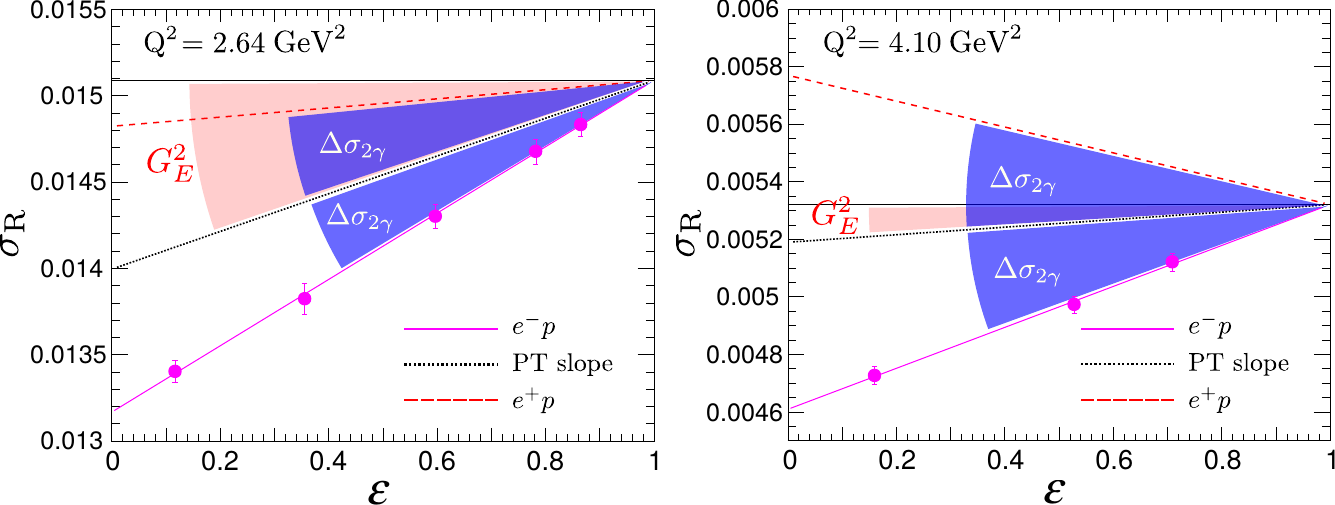}
    \caption{Reduced cross section from E01-001~\cite{Qattan:2004ht} (magenta points) along with linear fit. The solid black line is a constant at the $\varepsilon=1$ value of the cross section fit, while the dotted black line combined the $\varepsilon=1$ value from the fit and the slope based on polarization transfer measurements of $\mu_p G_E/G_M$. Neglecting TPE, the full difference solid black line and the magenta line at $\varepsilon=0$ yields $G_E^2$. Allowing for linear TPE corrections and assuming that the PT data is unaffected by TPE, the difference between the solid black line and the dashed black line represents $G_E^2$, while the difference between the solid black line and the magenta fit to the cross section represents the TPE contribution to the cross section, $\Delta \sigma_{2\gamma}$. The red dashed line represents the expected positron reduced cross section based on the linear TPE assumption described above, changing the sign of the $\Delta \sigma_{2\gamma}$ correction. The positron Super-Rosenbluth measurement should yield uncertainties as good or better than those of the E01-001 experiment. Figure adapted from Ref.~\cite{yurov2017}.}
    \label{fig:SR_pos_elec}
\end{figure}

To take advantage of these benefits, the Super-Rosenbluth measurement requires measurements at multiple beam energies to perform Rosenbluth separations at each $Q^2$ value measured. This allows for cancellation of many systematic uncertainties that are exactly (or nearly) identical for different $\varepsilon$ settings at fixed $Q^2$. Measurements with a larger number of energies improve the sensitivity to the $\varepsilon$ dependence which is especially beneficial when looking for non-linear contributions. Figure~\ref{fig:SR_pos_elec} shows the reduced cross section measurements from E01-001~\cite{Qattan:2004ht}, the slope expected based on polarization measurements (assuming no TPE corrections to the polarization observables), and a projection for the expected positron measurements. Electron measurements for $Q^2=4.1$~GeV$^2$ suggest that the contribution from $G_E(Q^2)$ is extremely small and that the slope extracted in the Rosenbluth separation mainly reflects TPE contributions, yielding a negative slope (unphysical in the one-photon exchange approximation, where the slope is proportional to ($G_E/G_M$)$^2$) for positron measurements.

\section{Proposed positron Super-Rosenbluth measurements}\label{sec:experiment}

Because the main benefit of the Super-Rosenbluth technique relies on cancellation between corrections at fixed $Q^2$ but different $\varepsilon$ values, rather than cancellation between corrections for positron and electron beams, the experiment can be performed with only positrons and compared to existing electron Super-Rosenbluth measurements. However, it is beneficial to make positron and electron measurements using the same detectors, as the resolution of the measured proton's angle and momentum is important in isolating elastic scattering and avoiding inelastic backgrounds~\cite{Qattan:2004ht}. Thus, the approach taken here is to optimize a set of positron SR measurements, and then to make the same measurements using electron beams. The positron current will be limited by the source, while the electron beams can be run at larger currents, such that the time is dominated by positron measurements. For the following projections, we assume a 2~$\mu$A positron beam current and use the existing SR measurements to make projections for statistical and systematic uncertainties.

The initial SR measurements~\cite{Qattan:2004ht} were performed in Hall A at Jefferson Lab~\cite{Alcorn:2004sb}, with an average beam current of 60~$\mu$A impinging on a 4~cm liquid hydrogen target, with an allocation of 10 days of beamtime. Precise extractions of $\mu G_E/G_M$ were made at $Q^2=$2.64, 3.2, and 4.1~GeV$^2$, with significantly smaller corrections and uncertainties than any other Rosenbluth separations in this kinematic region. Accounting for the reduction to 2~$\mu$A beam current for positrons and replacing the 4~cm target with a 10~cm target gives a measurement with a factor of 12 reduction in luminosity compared to the previous experiment. Because only one of the High Resolution Spectrometers (HRSs) was used for these measurements in the original experiment, we will make up a factor of two by using both spectrometers. We can save another factor of two by reducing the statistics since, even for the highest $Q^2$ setting, the statistical uncertainties were below the systematic uncertainties of the measurement, usually by a significant factor. In this scenario, we increase the run time by a factor of three, yielding a 30 day measurement that would provide nearly identical final uncertainties on the extracted value of $\mu G_E/G_M$ and slightly reduced sensitivity to deviations from linearity in the reduced cross section due to the slightly larger statistical uncertainties.

The Hall A measurement ran with five beam energies corresponding to two different linac energy settings. The follow-up measurement E05-017~\cite{E05017} ran in Hall C with 17 beam energies~\cite{yurov_phd}, allowing for a larger $\varepsilon$ range and more $\varepsilon$ points at each $Q^2$, with two dedicated linearity scans with 10 or more $\varepsilon$ points. The experiment used similar energies and target as the Hall A experiment and covered $Q^2$ values from 0.4-5.8~GeV$^2$ with 30 days of beamtime. A full version of this measurement using positrons is not feasible: as with the original measurement only the High-Momentum Spectrometer (HMS) can cover the necessary kinematic range, and the high-$Q^2$ points are statistics limited, meaning a significant reduction in statistics would significantly reduce the sensitivity. As such, one would have to make up the full factor of 12 in luminosity through increased run time. Thus, we base our projections on the Hall A electron measurements presented above. Note that while the plan presented here assumes data taking in Hall A with the two HRS spectrometers, the experiment could also be performed in Hall C with the HMS spectrometer with essentially the same figure of merit; Experiment E01-001 used the central 1.6~msr of the HRS, while E05-017 used 3.2~msr in the HMS.

\begin{figure}
    \centering
    \includegraphics[width=0.96\textwidth]{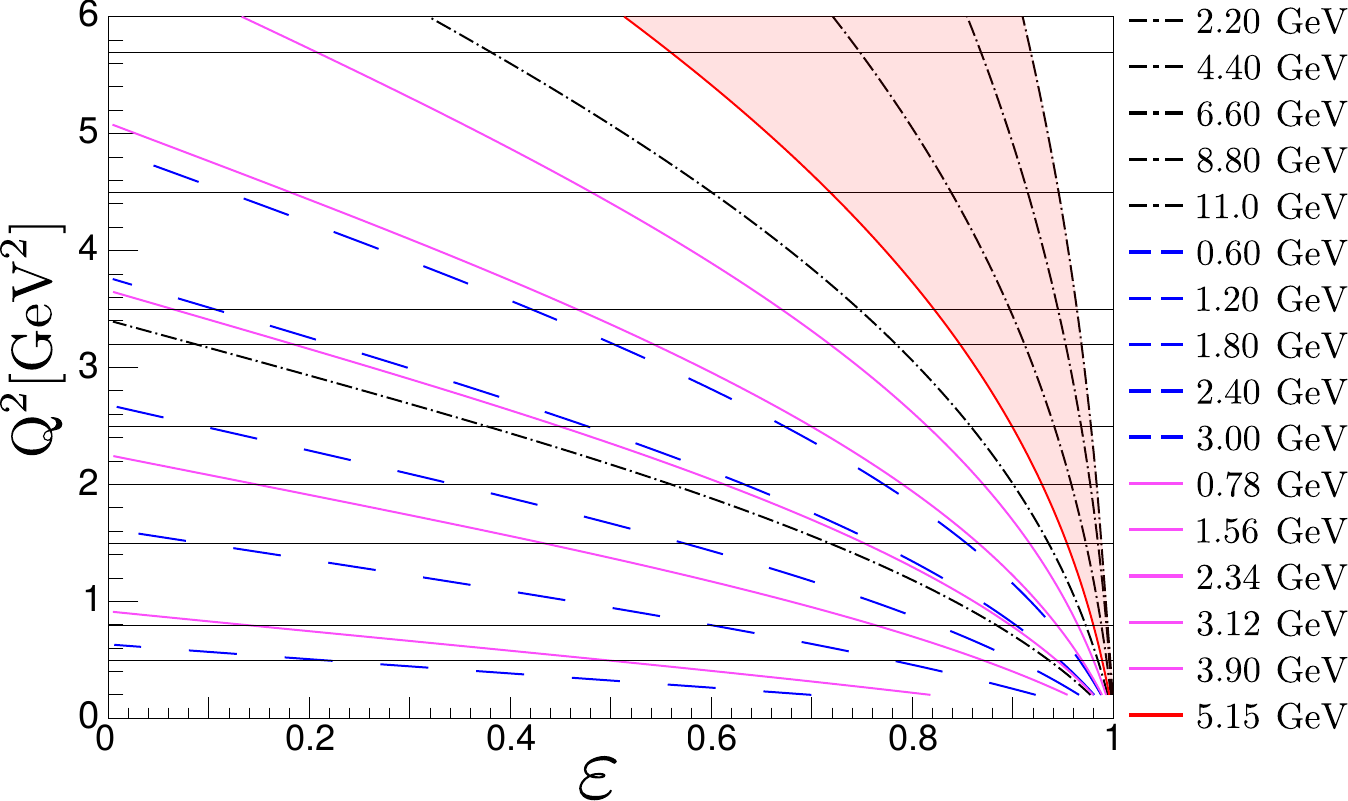}
    \caption{Potential kinematics for the proposed measurement. The curves indicate the elastic kinematics for beam energies corresponding to an energy per pass of 2.2 GeV (solid line), 0.78 GeV (short-dash), and 0.6 GeV. Horizontal lines represent $Q^2$ values that provide a good lever arm in $\varepsilon$. Measurements up to $Q^2 = 4.5$~GeV$^2$ are straightforward under the assumptions given in the text, and higher beam currents or a longer target would allow a precision measurements at $Q^2 \approx 5.7$~GeV$^2$. The red line indicates the highest beam energy used in previous measurements~\cite{E05017,yurov_phd}, and the red shaded region indicates the increased $\varepsilon$ coverage with higher energies. Above $Q^2 \approx 3$~GeV$^2$, the higher beam energies will provide a significant increase in the $\varepsilon$ coverage and a corresponding reduction in the uncertainty on $\mu_p G_E/G_M$.} 
    \label{fig:kinematics}
\end{figure}

Corresponding electron measurements could be taken with a factor of 10 or more increase in beam current, meaning that the electron measurements could be performed with minimal beam time for running, plus overhead for the required beam energy changes. While one could compare the positron SR separations to polarization measurements directly, as was done with the electron SR (illustrated in Fig.~\ref{fig:SR_pos_elec}), comparing electron and positron SR measurements doubles the size of the TPE effects observed in extracting the Rosenbluth slope or the deviations from linearity since the TPE contributions have the opposite sign for positrons and electrons. It also makes the comparison independent of TPE contributions to the polarization observables, although these are believed to be very small~\cite{Guichon:2003qm,Afanasev:2005ex,Meziane:2010xc}.

Note that the uncertainties in $R_p$ should match those from experiment E01-001~\cite{Qattan:2004ht} assuming measurements at identical kinematics. However, there is an additional gain that comes from the increased reach in $\varepsilon$ possible with an 11~GeV beam (compared to the 4.7~GeV (5.15~GeV) maximum beam energy from the Hall A (Hall C) measurement). This increases the lever arm in $\varepsilon$ by a factor of 1.5 for $Q^2=4.1$~GeV$^2$, reducing the extracted uncertainty in $R_p$ by an identical factor, making the positron measurement at least as precise as the completed electron measurement, or allowing for comparable precision at higher $Q^2$ values. Therefore, at the cost of additional overhead for beam energy changes, the $Q^2$ range could be increased somewhat while yielding precision identical to the previous measurement, and additional measurements could be added for several $Q^2$ values below 3~GeV$^2$, where the run times are minimal.

\begin{figure}
    \centering
    \includegraphics[width=0.96\textwidth]{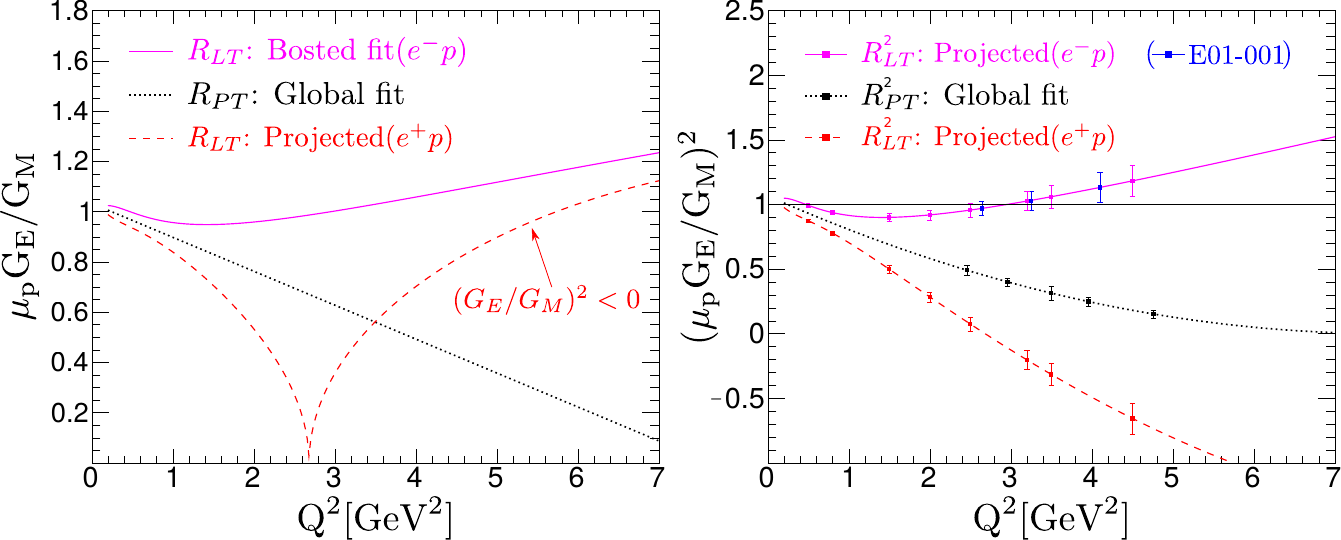}
    \caption{The left figure shows $\mu_pG_E/G_M$ from the Bosted fit to electron scattering data (top magenta curve), a parameterization of the polarization transfer results (black curve), and a prediction for the results of positron LT separations, assuming that TPE yields the difference. Note that for $Q^2 > 2.7$~GeV$^2$, the slope in the Rosenbluth separation for positrons becomes negative, yielding ($G_E/G_M)^2<0$ in the one-photon exchange approximation (as discussed at the end of Sec.~\ref{sec:SR-positron}). The right figure shows the same curves, but for $(\mu_pG_E/G_M)^2$. The blue and black points represent uncertainties on existing SR and polarization measurements, respectively (placed on the parameterizations), and the red and magenta point indicate the projected uncertainties for the proposed measurements.} 
    \label{fig:results}
\end{figure}

Figure~\ref{fig:kinematics} shows an example of the kinematic coverage that would be possible using three different values of the beam energy per pass. Note that these linac settings also allow for a measurement at 5.7~GeV$^2$ (not assumed in the scenario presented above), given additional time or running with higher luminosity. It would also allow for significantly improved checks on linearity, with more points and a wider range in $\varepsilon$ for $Q^2$ values up to 2-3~GeV$^2$. Changing from 3 $Q^2$ points from 2.64-4.1~GeV$^2$ to a total of 8 $Q^2$ values from 0.5-4.5~GeV$^2$, with additional $\varepsilon$ points for measurements below $Q^2=2.5$~GeV$^2$, would only increase the measurements by 3-5 days. Figure~\ref{fig:results} shows projections for the proposed measurements on positrons (and electrons), compared to a subset of polarization measurements and the E01-001~\cite{Qattan:2004ht} Super-Rosenbluth results.

\section{Conclusions}

In conclusion, we have presented a detailed plan for a Super-Rosenbluth measurement utilizing proposed positron beams at Jefferson Lab~\cite{Accardi:2020swt}. Based on the results of the previous Super-Rosenbluth measurement, we show that a 2~$\mu$A positron beam at Jefferson Lab would allow for a series of positron SR measurements over a wide range of $Q^2$ (0.5-4.5~GeV$^2$) and $\varepsilon$, covering the region where TPE are large and believed to explain the discrepancy between polarization and Rosenbluth extractions of the form factors. The measurement will provide improved precision compared to previous SR measurements, and will include electron SR measurements to be made at the same $Q^2$ values. This will more than double the sensitivity the TPE corrections believed to explain the discrepancy, as well as dedicated tests for non-linear contributions at larger $Q^2$, and extending to higher $\varepsilon$ values.

The existing electron Super-Rosenbluth measurement already provides the world's best precision on $\mu G_E/G_M$ from Rosenbluth experiments and the best constraints on the linear contribution of TPE based on comparisons to polarization measurements. They also provide the best limits on nonlinear TPE contribution. A direct comparison of electron and positrion Super-Rosenbluth measurements would double the sensitivity of these studies, while the 11 GeV beam energy expands the $Q^2$ and $\varepsilon$ range accessible.
Such a measurement would provide the first direct test of the hypothesis that TPE contributions explain the observed form factor discrepancy at $Q^2 > 1.5$~GeV$^2$, and would significantly increase our quantitative extraction of TPE contributions as a function of $Q^2$ and $\varepsilon$, validating form factor extractions and giving greater confidence in TPE calculations that must be applied in other reactions where direct, or even indirect, experimental constraints on TPE are not feasible.

\begin{acknowledgements}
This work was supported by U.S. Department of Energy, Office of Science, Office of Nuclear Physics, under contract number DE-AC02-05CH11231.

\end{acknowledgements}

% BibTeX users please use one of
%\bibliographystyle{spbasic}      % basic style, author-year citations
%\bibliographystyle{spmpsci}      % mathematics and physical sciences

\bibliographystyle{spphys}       % APS-like style for physics
\bibliography{references}   % name your BibTeX data base

\end{document}